\documentclass[11pt]{article}
\usepackage{amssymb}
\begin{document}
\setlength{\baselineskip}{15pt}
\title{Characterization, global analysis and integrability \\ of a family of Poisson structures}
\author{ \mbox{} \\ Benito Hern\'{a}ndez-Bermejo $^1$}
\date{}

\maketitle
\begin{center}
{\em Departamento de Matem\'{a}tica Aplicada. Universidad Rey Juan Carlos. \\
Calle Tulip\'{a}n S/N. 28933--M\'{o}stoles--Madrid. Spain.} 
\end{center}

\mbox{}

\mbox{}

\mbox{}

\begin{center} 
{\bf Abstract}
\end{center}
\noindent
An $n$-dimensional solution family of the Jacobi equations is characterized and investigated, including the global determination of its main features: the Casimir invariants, the construction of the Darboux canonical form and the proof of integrability for the related Poisson systems. Examples are given and include novel Poisson formulations.

\mbox{}

\mbox{}

\mbox{}

\mbox{}



\mbox{}


\vfill

\noindent $^1$ Telephone: (+34) 91 488 73 91. Fax: (+34) 91 488 73 38. \newline 
\mbox{} \hspace{0.05cm} E-mail: benito.hernandez@urjc.es 

\pagebreak
\begin{flushleft}
{\bf 1. Introduction}
\end{flushleft}

The use of finite-dimensional Poisson systems (see \cite{olv1} for an overview and a historical discussion) is ubiquitous in most fields of nonlinear dynamics, including domains such as mechanics, electromagnetism, mathematical biology, optics, etc. In fact, recasting a given vector field as a Poisson system allows the use of very diverse techniques and specific methods adapted to that format, embracing stability analysis, numerical integration, perturbation methods, or integability properties and characterization of invariants, just to mention a sample. For instance, see the discussions in \cite{bn1}-\cite{bn3} for a brief account of the successes of the finite-dimensional Poisson systems theory.

When expressed in coordinates, a dynamical system defined in a domain $\Omega \subset 
\mathbb{R}^n$ is said to be a finite-dimensional Poisson system if it can be written in the form
\begin{equation}
    \label{nd1nham}
    \frac{\mbox{d}x_i}{{\mbox{d}t}} = \sum_{j=1}^n J_{ij}(x) \partial _j H(x) 
	\; , \;\:\; i = 1, \ldots , n, 
\end{equation} 
or briefly $\dot{x}= {\cal J}(x) \cdot \nabla H(x)$, where $ \partial_j \equiv \partial / \partial x_j$ and function $H$, which is by construction a time-independent first integral, is the Hamiltonian. The structure functions $J_{ij}(x)$ are the entries of an $n \times n$ structure matrix ${\cal J}(x)$ and they are solutions of the Jacobi PDEs: 
\begin{equation}
     \label{nd1jac}
     \sum_{l=1}^n ( J_{li} \partial_l J_{jk} + J_{lj} \partial_l J_{ki} + 
     J_{lk} \partial_l J_{ij} ) = 0 \:\; , \;\:\;\: i,j,k=1, \ldots ,n
\end{equation}
The structure functions $J_{ij}(x)$ have to verify also the skew-symmetry condition:
\begin{equation}
     \label{nd1sksym}
     J_{ij} =  - J_{ji} \;\:\:\:\: \mbox{for all} \:\; i,j
\end{equation}

Several reasons justify the interest of Poisson dynamical systems. One is that they provide a wide generalization of classical Hamiltonian systems, allowing not only for odd-dimensional vector fields, but also because a structure matrix verifying 
(\ref{nd1jac}-\ref{nd1sksym}) admits a great diversity of forms apart from the classical (constant) symplectic matrices. Actually, Poisson systems are a generalization of the classical Hamiltonian systems on which a generalized bracket is defined, namely:
\begin{equation}
\label{nd1cp}
	\{ f(x),g(x) \} = \sum _{i,j=1}^n \partial_i f(x) J_{ij}(x) \partial_j g(x)
\end{equation}
for every pair of smooth functions $f(x)$ and $g(x)$. The possible rank degeneracy of the structure matrix ${\cal J}$ implies that a certain class of first integrals ($C(x)$ in what follows) termed Casimir invariants exist. There is no analog in the framework of classical Hamiltonian systems for such constants of motion, which are characterized by the property of having a null bracket ---in the sense of (\ref{nd1cp})--- with all smooth functions defined in $\Omega$. It can be seen that this implies that Casimir invariants are the solution set of the system of coupled PDEs: ${\cal J} \cdot \nabla C =0$. The determination of Casimir invariants and their use in order to carry out a reduction (local, in principle) is the cornerstone of the (at least local) equivalence between Poisson systems and classical Hamiltonian systems, as stated by Darboux' theorem 
\cite{olv1}. This justifies that Poisson systems can be regarded, to a large extent, as a rightful generalization of classical Hamiltonian systems. This connection is an additional and important advantage of Poisson systems, specially if the reduction can be accomplished globally. It is also clear that the issue of describing a given vector field not explicitly written in the form (\ref{nd1nham}) in terms of a Poisson structure is a fundamental question in this context, which still remains as an open problem (for instance, see 
\cite{5}-\cite{21} and references therein). This is a nontrivial decomposition to which important efforts have been devoted in past years in a variety of approaches. The source of the difficulty is obviously twofold: First, a known constant of motion of the system able to play the role of Hamiltonian is required. And second, it is necessary to find a suitable structure matrix for the vector field. Consequently, finding a solution of the Jacobi identities (\ref{nd1jac}) complying also with conditions (\ref{nd1sksym}) is unavoidable. This explains the intrinsic interest deserved in the literature by the obtainment and classification of skew-symmetric solutions of the Jacobi equations \cite{bn1}-\cite{bn3},\cite{tur}-\cite{29}. In this context, it is of special interest the characterization of $n$-dimensional solution families (with $n$ arbitrary) leading to Poisson systems for which the Darboux construction can be globally determined. The reason is twofold: in one hand, we have that while 
three and four-dimensional Poisson structures have been widely investigated (for instance, see 
\cite{bn1,bn2,byv1,tur,dhns,gyn1,bs1,bs2,nut1,27} and the discussions included in these works) the number of known families of arbitrary dimension is quite limited 
\cite{olv1,bn3,21,byr1,byv4,lyx1,pla1,29}. In second place, it seems desiderable to construct Darboux reductions globally valid in the domain of interest in order to fully connect those families of Poisson systems with their classical Hamiltonian formulation. In this sense, it is worth noting that the number of $n$-dimensional families for which an explicit and global construction of the Darboux canonical form has been provided is exceedingly limited 
\cite{bn3,byv4,byv2}. Again, this number increases slightly if we consider dimensionally limited situations, such as those of dimensions three \cite{olv1,bn2,dhns,bs2} or four 
\cite{bn1}.

In this contribution an $n$-dimensional family of skew-symmetric solutions of the Jacobi equations (\ref{nd1jac}-\ref{nd1sksym}) is investigated. Such family is remarkably general, in particular being defined in terms of functions of arbitrary nonlinearity. This explains that well-known Poisson structures and systems of applied interest now appear embraced as particular cases, as it will be seen. In addition, such unification is not only conceptual, since it is possible to explicitly determine features such as the Casimir invariants and the construction of the Darboux canonical form, thus allowing a unified analysis of seemingly unrelated systems which avoids a case-by-case approach. Additionally, the methods developed are valid globally in phase space, thus improving the scope (local, in principle) of Darboux' theorem. These results also imply the proof of both algebraic and Liouville integrability for these Poisson systems. Examples include the first nonstandard family of Poisson structures for Quasi-Polinomial systems, as well as novel Poisson structures for $n$-dimensional Lotka-Volterra systems. 

The structure of the work is the following. In Section 2 the determination and the global analysis of a family of skew-symmetric solutions of the Jacobi equations are presented. Some instances are considered in Section 3. 

\pagebreak
\begin{flushleft}
{\bf 2. The family of solutions and its global analysis}
\end{flushleft}

The first result to be presented is the following one:

\mbox{}

\noindent{\bf Theorem 1.} 
{\em Let  $\eta (x)$ and $\varphi_i(x_i)$, for $i=1, \ldots ,n$, be functions defined in a domain $\Omega \subset \mathbb{R}^n$, all of which are $C^{\infty}(\Omega)$ and nonvanishing in $\Omega$. In addition let 
\begin{equation}
\label{nd1psi}
	\psi_i(x_i) = a_i \exp \left( \int \frac{\mbox{d}x_i}{\varphi_i(x_i)} \right) 
	\: , \:\:\: i=1, \ldots ,n
\end{equation}
where $a _i \neq 0$, $i = 1, \ldots , n$, are arbitrary nonzero real constants, and every exponent in (\ref{nd1psi})
\begin{equation}
\label{nd1pso}
	\int \frac{\mbox{d}x_i}{\varphi_i(x_i)} \: , \:\:\: i=1, \ldots ,n
\end{equation}
denotes one primitive of $1/ \varphi_i(x_i)$. Finally, let the functions 
$\omega_{ij}(x_i,x_j)$ be defined as
\[
	\omega_{ij}(x_i,x_j) = \psi_i(x_i) - \psi_j(x_j) \: , \:\:\: i,j=1, \ldots ,n
\]
and assume that $\omega_{ij}(x_i,x_j)$ is nonvanishing in $\Omega$ at least for one pair $(i,j)$. Then ${\cal J}=(J_{ij})$ is a family of $n$-dimensional Poisson structures globally defined in $\Omega$, where }
\begin{equation}
\label{nd1sol1}
	J_{ij}(x)= \eta (x) \varphi_i(x_i) \varphi_j(x_j) \omega_{ij}(x_i,x_j)
	\:\: , \:\:\:\:\: i,j = 1, \ldots ,n
\end{equation}

\mbox{}

\noindent{\bf Proof.} Skew-symmetry is evident in (\ref{nd1sol1}). We then substitute ${\cal J}$ in (\ref{nd1sol1}) into the Jacobi identities (\ref{nd1jac}) and obtain after some algebra:
\[
     \sum_{l=1}^n ( J_{li} \partial_l J_{jk} + J_{lj} \partial_l J_{ki} + 
     J_{lk} \partial_l J_{ij} ) = \eta T_1 + \eta^2 T_2
\]
where $T_1$ and $T_2$ are the following terms, to be examined separately:
\[
	T_1 = \sum_{l=1}^n \varphi_i \varphi_j \varphi_k \varphi_l (\partial_l \eta)
	( \omega_{il} \omega_{jk} + \omega_{jl} \omega_{ki} + \omega_{kl} \omega_{ij} )
\]
\[
	T_2 = \sum_{l=1}^n \left\{ \varphi_i \varphi_l \omega_{il} \left[ \delta_{lj} 
	\varphi_j^{\prime} \varphi_k \omega_{jk} + \delta_{lk} \varphi_j \varphi_k^{\prime} 
	\omega_{jk}+\varphi_j \varphi_k \left( \delta_{lj}\frac{\psi_j}{\varphi_j}- 
	\delta_{lk}\frac{\psi_k}{\varphi_k} \right) \right] \right. +
\]
\[
	\varphi_j \varphi_l \omega_{jl} \left[ \delta_{lk} 
	\varphi_k^{\prime} \varphi_i \omega_{ki} + \delta_{li} \varphi_k \varphi_i^{\prime}
	\omega_{ki}+\varphi_k \varphi_i \left( \delta_{lk}\frac{\psi_k}{\varphi_k}- 
	\delta_{li}\frac{\psi_i}{\varphi_i} \right) \right] +
\]
\[
	\left. \varphi_k \varphi_l \omega_{kl} \left[ \delta_{li} 
	\varphi_i^{\prime} \varphi_j \omega_{ij} + \delta_{lj} \varphi_i \varphi_j^{\prime}
	\omega_{ij}+\varphi_i \varphi_j \left( \delta_{li}\frac{\psi_i}{\varphi_i}- 
	\delta_{lj}\frac{\psi_j}{\varphi_j} \right) \right] \right\}
\]
Regarding $T_1$, if every $\omega_{ij}$ is substituted by its expression $\omega_{ij}=\psi_i-\psi_j$ and the result is simplified, it is found that:
\[
	\omega_{il} \omega_{jk} + \omega_{jl} \omega_{ki} + \omega_{kl} \omega_{ij} = 0
\]
It is thus demonstrated that $T_1=0$. We proceed now with $T_2$: expanding its expression and cancelling out similar terms, after a suitable rearrangement we arrive at: 
\[
	T_2 =  \varphi_i \varphi_j \varphi_k [\omega_{ij}\psi_j-\omega_{ik}\psi_k
	+\omega_{jk}\psi_k-\omega_{ji}\psi_i+\omega_{ki}\psi_i-\omega_{kj}\psi_j] = 0
\]
Therefore it is also $T_2=0$ and the proof is complete. \hfill Q.E.D.

\mbox{}

Some brief but relevant comments must be provided at this point. In first place recall that, as indicated in the theorem, for every $i$ the primitive (\ref{nd1pso}) obtained from $\varphi_i(x_i)$ must be chosen to be one and the same for all the entries of ${\cal J}$. However, the specific choice is actually arbitrary. In this sense, notice that if different integration constants are selected, then the outcome is also a member of the solution family, this time with rescaled parameters $a_i$. Secondly, note that by construction the functions $\psi_i(x_i)$ and $\omega_{ij}(x_i,x_j)$ are $C^{\infty}(\Omega)$. In third place, it is worth observing that the definition (\ref{nd1psi}) allows an alternative expression for the family just characterized, namely ${\cal J}=(J_{ij})$ can also be written as
\begin{equation}
\label{nd1sol1b}
	J_{ij}(x)= \eta (x)\frac{\psi_i(x_i)\psi_j(x_j)}{\psi_i^{\prime}(x_i)\psi_j^{\prime}	(x_j)}\omega_{ij}(x_i,x_j) = 
	\eta (x)\frac{\psi_i(x_i)\psi_j(x_j)}{\psi_i^{\prime}(x_i)\psi_j^{\prime}(x_j)}
	[\psi_i(x_i) - \psi_j(x_j)] 
\end{equation}
for all $i,j = 1, \ldots ,n$, where functions $\psi_i(x_i)$ must be $C^{\infty}(\Omega)$ and nonvanishing in $\Omega$, and such that functions $\psi_i^{\prime}(x_i)$ are also nonvanishing in $\Omega$, while the rest of defining properties were already presented in Theorem 1. Under these assumptions, (\ref{nd1sol1b}) can be taken as an alternative definition of the solution family of Poisson structures. 

Some of the properties of the family identified in Theorem 1 can be characterized now: 

\mbox{}

\noindent {\bf Theorem 2.} 
{\em Let ${\cal J}$ be a Poisson structure of the form (\ref{nd1sol1}) characterized in Theorem 1, which is defined in a domain $\Omega \subset \mathbb{R}^n$ and such that the pair $(i,j)$ verifies that function $\omega_{ij}(x_i,x_j)$ is nonvanishing in $\Omega$. 
Then Rank(${\cal J}$)$=2$ everywhere in $\Omega$ and a complete set of independent Casimir invariants for ${\cal J}$ is given by:}
\begin{equation}
\label{nd1cas}
	C_{k}(x) = \frac{\psi_i(x_i)[\psi_j(x_j) - \psi_k(x_k)]}{\psi_k(x_k)[\psi_i(x_i) - 
	\psi_j(x_j)]} = \frac{\psi_i(x_i)\omega_{jk}(x_j,x_k)}{\psi_k(x_k)\omega_{ij}(x_i,x_j)} 
	\: , \:\:\:\:\: k = 1, \ldots ,n \: ; \:\:\: k \neq i,j
\end{equation}
{\em Moreover, every Casimir invariant in (\ref{nd1cas}) is globally defined in $\Omega$ and $C^{\infty}(\Omega)$.}

\mbox{} 

\noindent{\bf Proof.} Since functions $\eta(x)$ and $\varphi_i(x_i)$ are nonvanishing in $\Omega$, the use of rank-preserving matrix transformations shows that Rank(${\cal J}$) $=$ Rank(${\cal W}$) in $\Omega$, where ${\cal W} \equiv (\omega_{ij}(x_i,x_j))$. Since at least one of the entries of ${\cal W}$ is also nonvanishing in $\Omega$, this implies that Rank(${\cal J}$) $\geq 2$ everywhere in $\Omega$. We can now submit matrix ${\cal W}$ to additional rank-preserving transformations: notice that Rank(${\cal W}$) is also maintained if we substract the first row to the rest of rows, and then if on the resulting matrix we substract the first column to every one of the remaining columns. This leads to a new matrix ${\cal W}^*$ given by:
\begin{equation}
\label{nd1xbis}
	{\cal W}^* = \left( \begin{array}{cccc}
				0     &  \omega_{12}  &  \ldots  &  \omega_{1n}  \\
			- \omega_{12} &       0     &  \ldots  &     0       \\
			  \vdots    &   \vdots    &  \ddots  &  \vdots     \\
			- \omega_{1n} &       0     &  \ldots  &     0       \\
			 \end{array} \right)
\end{equation}
Then, it is clear from (\ref{nd1xbis}) that Rank(${\cal J}$) $=$ Rank(${\cal W}^*$) $\leq 2$ in every point of $\Omega$. Therefore we conclude that Rank(${\cal J}$) $=2$ in $\Omega$. This demonstrates the first part of the statement. For the second part, notice first that every function $C_k(x)$ in (\ref{nd1cas}) always depends on $x_i$, $x_j$ and $x_k$ (since functions $\psi_k(x_k)$ cannot be constant for any $k$, according to the conditions established) and in addition $C_k(x)$ does not depend on the rest of variables. This implies immediately the functional independence of the set $\{ C_k(x): k= 1, \ldots ,n; k \neq i,j \}$. Moreover, since all functions composing $C_k(x)$ are $C^{\infty}(\Omega)$ and $\psi_k(x_k)\omega_{ij}(x_i,x_j) \neq 0$ everywhere in $\Omega$, function $C_k(x)$ is necessarily $C^{\infty}(\Omega)$. Therefore, to complete the proof it is only required to demonstrate that functions $C_k(x)$ are Casimir invariants for every $k$. The most simple way to see this is to verify that ${\cal J} \cdot \nabla C_k=0$ for every $k=1, \ldots , n$, with $k \neq i,j$ (notice that for both values $k=i,j$, function $C_k(x)$ is a constant, and then also a Casimir invariant, but trivial). We thus have: 
\[
	\partial _i C_k(x) = \frac{\psi_i^{\prime} \psi_j \psi_k \omega_{kj}}{
	( \psi_k \omega_{ij})^2} \: , \:\;\:
	\partial _j C_k(x) = \frac{\psi_i \psi_j^{\prime} \psi_k \omega_{ik}}{
	( \psi_k \omega_{ij})^2} \: , \:\;\:
	\partial _k C_k(x) = \frac{\psi_i \psi_j \psi_k^{\prime} \omega_{ji}}{
	( \psi_k \omega_{ij})^2} 
\]
for $k=1, \ldots n$, $k \neq i,j$. Then for every $r=1, \ldots ,n$ it can be seen that:
\[
	\sum_{s=1}^n J_{rs} \partial _s C_k = J_{ri} \partial_i C_k + J_{rj} \partial_j C_k +
	J_{rk} \partial_k C_k = 
\]
\begin{equation}
\label{nd1t2jnc}
	\frac{\eta \varphi_r \psi_i \psi_j }{\psi_k (\omega_{ij})^2}
	(\omega_{ri}\omega_{kj} + \omega_{rj}\omega_{ik}+ \omega_{rk}\omega_{ji})
\end{equation}
In equation (\ref{nd1t2jnc}) the last term vanishes for every choice of $i,j,k,r$,
\[
	\omega_{ri}\omega_{kj} + \omega_{rj}\omega_{ik}+ \omega_{rk}\omega_{ji} = 0
\]
as it was already shown in the proof of Theorem 1. Consequently, ${\cal J} \cdot \nabla C_k =0$ for every $k$. This completes the proof. \hfill Q.E.D.

\mbox{}

Consequently, every Poisson system of this kind has $(n-2)$ independent Casimir invariants, additional to the Hamiltonian. In other words:

\mbox{}

\noindent {\bf Corollary 1.} 
{\em Every Poisson system $\dot{x}= {\cal J}(x) \cdot \nabla H(x)$ in which the structure matrix ${\cal J}(x)$ is of the kind specified in Theorem 1 is an algebraically integrable system.}

\mbox{}

Before proceeding further, it is necessary to recall the concept of time reparametrization for Poisson systems \cite{gyn1,bs2}, which are transformations of the form
\begin{equation}
\label{nd1ntt}
	\mbox{d}\tau = \frac{1}{\mu (x)}\mbox{d}t
\end{equation}
where $t$ is the initial time variable, $\tau$ is the new time and $\mu (x) : \Omega 
\rightarrow \mathbb{R}$ is a $C^{\infty}(\Omega)$ function which does not vanish in $\Omega$. Thus, if 
\begin{equation}
\label{nd13dpos}
	\frac{\mbox{d}x}{\mbox{d}t} = {\cal J} \cdot \nabla H
\end{equation}
is an arbitrary Poisson system defined in $\Omega$, then every time reparametrization 
(\ref{nd1ntt}) leads from (\ref{nd13dpos}) to the system (not necessarily of Poisson type):
\begin{equation}
\label{nd13dposntt}
	\frac{\mbox{d}x}{\mbox{d} \tau} = \mu {\cal J} \cdot \nabla H
\end{equation}
Having this issue in mind, an additional consequence of the previous results is that they allow the case-classification for the constructive and global determination of the Darboux canonical form for this kind of Poisson systems. This statement is contained in the following:

\mbox{}

\noindent {\bf Theorem 3.} 
{\em Let $\Omega \subset \mathbb{R}^n$ be a domain where a Poisson system (\ref{nd1nham}) is defined everywhere, for which ${\cal J}$ is a structure matrix of the form (\ref{nd1sol1}) characterized in Theorem 1, and such that the pair $(i,j)$ verifies that function $\omega_{ij}(x_i,x_j)$ is nonvanishing in $\Omega$. Then such Poisson system can be globally reduced in $\Omega$ to a one degree of freedom Hamiltonian system and the Darboux canonical form is accomplished globally in $\Omega$ in the new coordinate system $(y_1, \ldots ,y_n)$ and the new time $\tau$, where $(y_1, \ldots ,y_n)$ are given by the diffeomorphism globally defined in $\Omega$ 
\begin{equation}
\label{nd1darbco}
	\left\{ \begin{array}{ccl}
	y_i (x) & = & x_i  \\
	y_j (x) & = & x_j  \\
	y_k (x) & = & C_k(x) \;\: , \;\:\;\: k=1, \ldots ,n ; \:\:\; k \neq i,j
	\end{array} \right.
\end{equation}
in which the $C_k(x)$ are the Casimir invariants (\ref{nd1cas}); and the new time $\tau$ is defined by the time reparametrization:
}
\begin{equation}
\label{nd1darbntt}
	\mbox{d} \tau = J_{ij}(x(y)) \mbox{d} t
\end{equation}

\mbox{}

\noindent{\bf Proof.} According to Theorem 2, Darboux' theorem is applicable because ${\cal J}$ has constant rank 2 in $\Omega$. For the sake of clarity and without loss of generality, assume that it is $\omega_{12} \neq 0$ everywhere in $\Omega$. Recall also that, after a general diffeomorphism $y = y(x)$, an arbitrary structure matrix ${\cal J}(x)$ is transformed into another one ${\cal J^*}(y)$ as:
\begin{equation}
\label{nd1jdiff}
      J^*_{ij}(y) = \sum_{k,l=1}^n \partial _k y_i \partial _l y_j J_{kl}(x)  
	\;\; , \;\:\; i,j = 1, \ldots , n
\end{equation}
For the family of interest, the reduction is carried out in two steps. We first perform the change of variables 
(\ref{nd1darbco}), which in this case is
\begin{equation}
\label{nd1d12}
	\left\{ \begin{array}{ccl}
	y_1 (x) & = & x_1  \\
	y_2 (x) & = & x_2  \\
	y_k (x) & = & C_k(x) \;\: , \;\:\;\: k=3, \ldots ,n 
	\end{array} \right.
\end{equation}
where the $C_k(x)$ are given by (\ref{nd1cas}), namely:
\begin{equation}
\label{nd1casdb}
	C_k(x) = \frac{\psi_1(x_1)\omega_{2k}(x_2,x_k)}{\psi_k(x_k)\omega_{12}(x_1,x_2)}
	= \frac{\psi_1(x_1)(\psi_2(x_2)- \psi_k(x_k))}{\psi_k(x_k)(\psi_1(x_1) - \psi_2(x_2))}
	\;\: , \;\:\;\: k=3, \ldots ,n 
\end{equation}
Note that this change of variables is invertible everywhere in $\Omega$, its inverse being:
\begin{equation}
\label{nd1invd12}
	\left\{ \begin{array}{ccl}
	x_1 (y) & = & y_1  \\
	x_2 (y) & = & y_2  \\
	x_k (y) & = & \displaystyle{ \zeta_k\left[ \frac{\psi_1(y_1)\psi_2(y_2)}{\psi_1(y_1)+
		y_k \omega_{12}(y_1,y_2)} \right] } \;\: , \;\:\;\: k=3, \ldots ,n 
	\end{array} \right. 
\end{equation}
where function $\zeta_k$ is the inverse function of $\psi_k$ for every $k$. The examination of (\ref{nd1d12}-\ref{nd1invd12}) shows that the variable transformation (\ref{nd1d12}) to be performed exists and is a diffeomorphism everywhere in $\Omega$ as a consequence that by hypothesis we have $\omega_{12}(x_1,x_2) \neq 0$ in $\Omega$, as well as $\psi_k(x_k) \neq 0$ and $\psi^{\prime}_k(x_k) \neq 0$ for every $k$ in $\Omega$. Then, according to (\ref{nd1d12}) and (\ref{nd1casdb}), and taking (\ref{nd1jdiff}) into account, after some algebra we are led to
\begin{equation}
\label{nd1jdarb1}
	{\cal J^*}(y) = J_{12}(x(y)) \left( \begin{array}{ccccc}
		 0 & 1 & 0 & \ldots & 0 \\ 
		-1 & 0 & 0 & \ldots & 0 \\ 
		 0 & 0 & 0 & \ldots & 0 \\ 
		 \vdots & \vdots & \vdots & \ddots & \vdots \\
		 0 & 0 & 0 & \ldots & 0
		\end{array} \right)
\end{equation}
where from equations (\ref{nd1sol1}) and (\ref{nd1invd12}) we have
\begin{equation}
\label{nd1j12ntt}
	J_{12}(x(y)) = \eta (y_1,y_2,x_3(y), \ldots , x_n(y)) \varphi_1(y_1) \varphi_2(y_2) 
				\omega_{12}(y_1,y_2)
\end{equation}
The explicit dependences of $( x_3(y), \ldots , x_n(y) )$ are obviously the ones given in (\ref{nd1invd12}) and were not displayed in (\ref{nd1j12ntt}) for the sake of clarity. Note that $J_{12}(x(y))$ is nonvanishing in $\Omega ^* = y(\Omega)$ and $C^{\infty}(\Omega ^*)$. These properties allow the accomplishment of the second step of the reduction which is a reparametrization of time, which in this case does not suppress the Poisson structure of the vector field. Thus, making use of (\ref{nd1j12ntt}) in equation (\ref{nd1darbntt}), the transformation $\mbox{d} \tau = J_{12}(x(y)) \mbox{d} t$ is performed. According to (\ref{nd1ntt}-\ref{nd13dposntt}) this leads from the structure matrix (\ref{nd1jdarb1}) to the Darboux canonical one:
\begin{equation}
\label{nd1jdarb}
	{\cal J}_D (y) = \left( \begin{array}{ccccc}
		 0 & 1 & 0 & \ldots & 0 \\ 
		-1 & 0 & 0 & \ldots & 0 \\ 
		 0 & 0 & 0 & \ldots & 0 \\ 
		 \vdots & \vdots & \vdots & \ddots & \vdots \\
		 0 & 0 & 0 & \ldots & 0
		\end{array} \right)
\end{equation}
Now the reduction is globally completed.  \hfill Q.E.D.

\mbox{}

The analysis of the family of Poisson structures is thus concluded, since at this stage the reduction directly connects the initial Poisson systems with their classical Hamiltonian formulations. In particular, we have:

\mbox{}

\noindent {\bf Corollary 2.} 
{\em Every Poisson system $\dot{x}= {\cal J}(x) \cdot \nabla H(x)$ in which the structure matrix ${\cal J}(x)$ is of the kind specified in Theorem 1 can be reduced globally and diffeomorphically to a Liouville integrable Hamiltonian system.}

\mbox{}

In what follows, the results just developed are illustrated by means of some examples. These provide several physical instances of the previous results, and also illustrate the procedures introduced. 

\pagebreak
\begin{flushleft}
{\bf 3. Examples}
\end{flushleft}

\noindent{\bf Example 1.} {\em Lotka-Volterra equations.}

\mbox{}

Let us first consider a Poisson structure employed in the analysis of the 3-d Lotka-Volterra (LV) equations \cite{gyn1}. The following LV system has deserved some attention in the literature \cite{gyn1,nut1,byv2},
\begin{equation}
\label{mpag1lv3}
	\left\{ \begin{array}{rcl}
	\dot{x}_1 & = & x_1 (\lambda_1 + a_2 x_2 + x_3) \\
	\dot{x}_2 & = &  x_2 (\lambda_2 + x_1 + a_3 x_3) \\
	\dot{x}_3 & = & x_3 (\lambda_3 + a_1 x_1 + x_2) 
	\end{array} \right.
\end{equation}
in which $x_i>0$ for all $i=1,2,3$. Among several classical integrable cases of interest, the following one is to be considered \cite{gyn1}:
\begin{equation}
\label{mpag1lv4}
	a_i = 1 \;\; , \;\; \lambda_i = 0 \;\; , \;\;\;\; i=1,2,3
\end{equation}
System (\ref{mpag1lv3}-\ref{mpag1lv4}) is Poisson, in terms of the structure matrix:
\begin{equation}
\label{mpaghalp1}
	J_{ij}(x) = x_i x_j (x_i - x_j) \;\; , \;\;\;\; i,j=1,2,3
\end{equation}
And the following first integral plays the role of Hamiltonian, 
\begin{equation}
\label{mpag1hlv}
	H(x)= \log \left[ \left( \frac{x_3}{x_1 x_2}(x_1 - x_2)^2 \right)^{-k} \left( 
	\frac{x_1}{x_2 x_3}(x_2 - x_3)^2 \right)^{k-1} \right] 
\end{equation}
for arbitrary $k \in \mathbb{R}$. It can be seen that the structure matrix (\ref{mpaghalp1}) belongs to the family (\ref{nd1sol1}) with $\eta (x) =1$, $\varphi_i(x_i)=x_i$ and $\psi_i(x_i)=x_i$ for all $i=1,2,3$. Since $x_i>0$ for all $i$, such structure is defined in every domain $\Omega \subset \mathbb{R}^3_+$. In addition, now $\omega _{ij}(x_i,x_j)=(x_i-x_j)$ for every pair $(i,j)$. Thus if $x_i \neq x_j$ in $\Omega$ for a pair $i \neq j$, then $\omega _{ij}(x_i,x_j) \neq 0$ (and therefore $J_{ij}(x) \neq 0$) in $\Omega$. Depending on $i$ and $j$, we have to employ according to (\ref{nd1cas}) different forms for the Casimir invariant. For instance, if $\omega_{12}(x_1,x_2) \neq 0$ in $\Omega$, we have:
\begin{equation}
\label{mpahcas}
	C_3(x) = \frac{x_1(x_2-x_3)}{x_3(x_1-x_2)}
\end{equation}
Therefore the reduction to Darboux form now makes use of the following diffeomorphism
\begin{equation}
\label{mpahdiff}
	y_1 = x_1 \;\: , \;\:\;\:
	y_2 = x_2 \;\: , \;\:\;\:
	y_3 = C_3(x) 
\end{equation}
with $C_3(x)$ given by (\ref{mpahcas}). The inverse of this transformation is then: 
\begin{equation}
\label{mpag1iex1}
	x_1 = y_1 \;\: , \;\:\;\:
	x_2 = y_2 \;\: , \;\:\;\:
	x_3 = \frac{y_1y_2}{y_1 + (y_1 - y_2) y_3}
\end{equation}
Notice that $y_1 + (y_1 - y_2) y_3=x_1x_2/x_3$ and consequently does not vanish, as expected. Thus, after applying (\ref{nd1jdiff}) the outcome is that ${\cal J}(x)$ in (\ref{mpaghalp1}) is transformed into:
\[
	{\cal J^*}(y) = y_1y_2(y_1 - y_2) \left( \begin{array}{ccc}
	0 & 1 & 0 \\ -1 & 0 & 0 \\ 0 & 0 & 0 \end{array} \right) \equiv
	\tilde{J}_{12}(y) \left( \begin{array}{ccc}
	0 & 1 & 0 \\ -1 & 0 & 0 \\ 0 & 0 & 0  \end{array} \right)
\]
The reduction is then completed by means of the time reparametrization $\mbox{d} \tau = \tilde{J}_{12}(y) \mbox{d} t$, which finally leads to the Darboux canonical form (\ref{nd1jdarb}) with $y_3$ acting as the decoupled Casimir and $(y_1,y_2)$ as canonical Hamiltonian variables. 

\mbox{}

\noindent {\bf Example 2.} {\em A nonstandard Quasi-Polynomial generalization of the 
Lotka-Volterra system.}

\mbox{}

In this second example the previous LV system is generalized as a Quasi-Polynomial (QP) flow in such a way that its associated Poisson structure is also generalized, while remaining in the framework of the family characterized in Theorem 1. The reader is referred to 
\cite{br1}-\cite{bv3} and references therein for an introduction to QP systems and their related formalism.  

Let us thus consider system (\ref{mpag1lv3}-\ref{mpag1lv4}) and perform the quasimonomial transformation 
\begin{equation}
\label{mpag1qmt}
	x_i=y_i^{c_i} \; , \;\; i=1,2,3 \; ; \:\:\;\; c_1c_2c_3 \neq 0 
\end{equation}
followed by the time reparametrization 
\begin{equation}
\label{mpag1nte}
	\mbox{d} \tau = \left( \prod_{i=1}^{3}c_i y_i^{c_i-1} \right)^{-1} \mbox{d} t 
\end{equation}
The outcome is the following QP generalization of the LV flow defined in terms of variables $y_i$ and time $\tau$:
\begin{equation}
\label{mpag1qp3}
	\left\{ \begin{array}{rcl}
	\dot{y}_1 & = & c_2 c_3 y_1^{c_1}y_2^{c_2-1}y_3^{c_3-1} (y_2^{c_2}+y_3^{c_3}) \\
	\dot{y}_2 & = & c_1 c_3 y_1^{c_1-1}y_2^{c_2}y_3^{c_3-1} (y_1^{c_1}+y_3^{c_3}) \\
	\dot{y}_3 & = & c_1 c_2 y_1^{c_1-1}y_2^{c_2-1}y_3^{c_3} (y_1^{c_1}+y_2^{c_2}) 
	\end{array} \right.
\end{equation}
Both transformations (\ref{mpag1qmt}-\ref{mpag1nte}) become identical in the case $c_1=c_2=c_3=1$, and system (\ref{mpag1qp3}) is thus reduced to (\ref{mpag1lv3}-\ref{mpag1lv4}) in such situation. On the other hand, equations (\ref{mpag1qp3}) are a Poisson system. Actually, the Hamiltonian (\ref{mpag1hlv}) is directly generalized as:
\[
	H^*(y)= \log \left[ \left( \frac{y_3^{c_3}}{y_1^{c_1}y_2^{c_2}}(y_1^{c_1} - y_2^{c_2})^2 
	\right)^{-k} \left( \frac{y_1^{c_1}}{y_2^{c_2} y_3^{c_3}}(y_2^{c_2} - y_3^{c_3})^2 
	\right)^{k-1} \right] 
\]
for arbitrary $k \in \mathbb{R}$. Finally, both the quasimonomial transformation and the time reparametrization transform the structure matrix (\ref{mpaghalp1}) leading to the more general form:
\begin{equation}
\label{mpaghalpx}
	J^*_{ij}(y) = y_i^{c_i} y_j^{c_j} (y_i^{c_i} - y_j^{c_j}) 
	\sum_{k=1}^3 (\epsilon _{ijk})^2 c_k y_k^{c_k-1} \; , \;\; i,j=1,2,3
\end{equation}
Structure matrix (\ref{mpaghalpx}) belongs to family (\ref{nd1sol1}) with $\eta (y) = c_1 c_2 c_3 y_1^{c_1-1}y_2^{c_2-1}y_3^{c_3-1}$, $\varphi_i(y_i)=y_i/c_i$ and $\psi_i (y_i) = y_i^{c_i}$ for $i=1,2,3$. Since we have $y_i>0$ for all $i$, (\ref{mpaghalpx}) is correctly defined in $\mathbb{R}_+^3$ without further assumptions. Now the reduction to the Darboux canonical form can be also performed globally and it is a generalization of the one for the LV case. For this, note first that if $(i,j,k)$ is a cyclic permutation of $(1,2,3)$ and $\omega _{ij}=(y_i^{c_i}-y_j^{c_j}) \neq 0$ in $\Omega$, then the Casimir invariant $C_k (y)$ is: 
\[
	C_k(y) = \frac{y_i^{c_i}(y_j^{c_j}-y_k^{c_k})}{y_k^{c_k}(y_i^{c_i}-y_j^{c_j})}
\]
Then, use of the corresponding transformation can be made in order to carry out the reduction to Darboux form. For instance, if $\omega _{12} \neq 0$:
\[
	z_1 = y_1 \;\: , \;\:\;\:
	z_2 = y_2 \;\: , \;\:\;\:
	z_3 = \frac{y_1^{c_1}(y_2^{c_2}-y_3^{c_3})}{y_3^{c_3}(y_1^{c_1}-y_2^{c_2})}
\]
And the inverse of this transformation is also a generalization of (\ref{mpag1iex1}): 
\[
	y_1 = z_1 \;\: , \;\:\;\:
	y_2 = z_2 \;\: , \;\:\;\:
	y_3 = \left( \frac{z_1^{c_1}z_2^{c_2}}{z_1^{c_1} + (z_1^{c_1} - z_2^{c_2}) z_3} 
	\right)^{1/c_3}
\]
The remaining details of the Darboux reduction are essentially similar to those of the LV case and therefore they are not given here. To conclude this example, it is worth noticing that extensive families of QP Poisson systems have been analyzed in detail in the literature 
\cite{byv2,jmaa}, but always in terms of a standard (in fact separable \cite{byv4} and quadratic) kind of Poisson structure. In such context the structure matrices (\ref{mpaghalpx}) are significant as far as they provide the first example known (to the author's knowledge) of a class of nonstandard Poisson structures associated with a family of QP systems.

\mbox{}

\noindent {\bf Example 3.} {\em Poisson structure for the system of circle maps.}

\mbox{}

As a next example the following Poisson structure, which is of interest for the analysis of the system of circle maps \cite{gyn1}, will be considered:
\begin{equation}
\label{mpajtop}
	J_{ij}(x) = \eta(x) x_i x_j (x_i - x_j) \;\; , \;\;\;\; i,j=1,2,3
\end{equation}
where
\begin{equation}
\label{mpajtop2}
	\eta (x) = -[(x_1-x_2)(x_2-x_3)(x_3-x_1)]^{-1}
\end{equation}
This structure matrix is to a great extent similar to the one in Example 1, apart from the factor $\eta (x)$ which nevertheless introduces some differences. As before, we have $\psi_i(x_i)=x_i$ and $\varphi_i(x_i)=x_i$ for every $i=1,2,3$. But according to Theorem 1, now the structure is defined provided that in $\Omega$ we have $x_i \neq 0$ for every $i$, and $x_i - x_j \neq 0$ for every pair $i \neq j$. If this is the case, function $\eta (x)$ is $C^{\infty} ( \Omega )$ and nonvanishing in $\Omega$. Note that the same conditions also imply $\omega _{ij}(x_i,x_j) \neq 0$ (and $J_{ij}(x) \neq 0$) in $\Omega$ for every pair $i \neq j$. Consequently, Theorem 2 implies that now every alternative form (\ref{nd1cas}) of the Casimir invariant is simultaneously defined in $\Omega$, namely: 
\[
	C_{1}(x) = \frac{x_2(x_3 - x_1)}{x_1(x_2 - x_3)} \: , \;\; 
	C_{2}(x) = \frac{x_3(x_1 - x_2)}{x_2(x_3 - x_1)} \: , \;\; 
	C_{3}(x) = \frac{x_1(x_2 - x_3)}{x_3(x_1 - x_2)} 
\]
Therefore, in order to perform the Darboux reduction of (\ref{mpajtop}-\ref{mpajtop2}) either expression can be employed. For instance, if we focus again on $C_3(x)$, which coincides with (\ref{mpahcas}), then transformation (\ref{mpahdiff}) is also the same. The rest of the reduction is thus analogous to the one in the Lotka-Volterra case, just with minor differences due to the presence of $\eta (x)$ as given by (\ref{mpajtop2}). Since such reduction does not present any feature not mentioned in the proof of Theorem 3, the rest will be omitted for the sake of conciseness. 

\mbox{}

\noindent {\bf Example 4.} {\em Poisson structures for $n$-dimensional Lotka-Volterra equations.}

\mbox{}

In order to provide an $n$-dimensional illustration of the results developed, we shall consider quadratic (namely homogeneous) Lotka-Volterra systems. Such models have already deserved some attention in the Poisson structure context \cite{pla1,pla3,jnla} in cases different from the one to be regarded here, but also in the $n$-dimensional framework. However, the possibility now considered seems to be new in the literature. Quadratic LV equations have the generic form
\begin{equation}
\label{qlv}
	\dot{x}_i = x_i \left( \sum_{j=1}^n \alpha_{ij}x_j \right) \;\: , \;\:\;\: i = 1 , 
		\ldots ,n
\end{equation}
with $x_i >0$ and $\alpha_{ij} \in \mathbb{R}$ for all $i,j=1, \ldots ,n$. In what follows we shall be concerned with the case (integrable, as it will be shown) in which
\begin{equation}
\label{parqlv}
	\alpha_{ij} = \left\{ \begin{array}{ll}
	\displaystyle{ a_i \sum_{\stackrel{\scriptstyle k=1}{\scriptstyle k \neq i}}^n b_k } & , 		\:\; i=j  \\
	\mbox{} & \mbox{} \\ 
	\displaystyle{ - a_jb_j } & , \:\; i \neq j
	\end{array} \right.
\end{equation}
where $a_i$ and $b_i$ are real constants, with $a_i \neq 0$, for $i=1, \ldots , n$. Equations (\ref{qlv}-\ref{parqlv}) obviously embrace both competitive and cooperative dynamical situations. Following Theorem 1, consider an open domain $\Omega \subset \mathbb{R}^n_+$ in which the Poisson structure is to be defined. Moreover, we set $\eta (x) = 1$, $\varphi_i(x_i)=x_i$ and consistently $\psi_i(x_i)=a_ix_i$ for every $i=1, \ldots ,n$. This leads to a natural 
$n$-dimensional generalization of the previous structure matrices (\ref{mpaghalp1}):
\begin{equation}
\label{nd1jhalpn}
	J_{ij}(x) = x_i x_j (a_ix_i - a_jx_j) \;\: , \;\:\;\: i,j=1, \ldots ,n
\end{equation}
The structure matrices (\ref{nd1jhalpn}) correspond to quadratic LV systems of the kind 
(\ref{qlv}-\ref{parqlv}) for Hamiltonian functions of the form:
\[
	H(x) = \sum_{i=1}^n b_i \log x_i
\]
In order to fully comply with the requirements of Theorem 1 (and necessarily for the application of Theorems 2 and 3) it must be also assumed that there exists at least one pair of indexes $(i,j)$ for which $\omega_{ij}(x_i,x_j)=a_ix_i-a_jx_j \neq 0$ everywhere in $\Omega$. Consistently with the previous style, in what follows this will be the case for $\omega_{12}$. Therefore, according to (\ref{nd1cas}) and Theorem 2, a complete set of $C^{\infty}( \Omega )$ and functionally independent Casimir invariants associated with the Poisson structures 
(\ref{nd1jhalpn}) are:
\begin{equation}
\label{nd1hcas}
	C_k(x) = \frac{a_1x_1(a_2x_2-a_kx_k)}{a_kx_k(a_1x_1-a_2x_2)} 
		\;\: , \;\:\;\: k=3, \ldots ,n
\end{equation}
Then the reduction to Darboux form now makes use of the diffeomorphism (\ref{nd1d12}), with 
the $C_k(x)$ given by (\ref{nd1hcas}). Since the reduction to the Darboux canonical form is a natural generalization of the one regarded in Example 1, the details are not provided here for brevity.

\pagebreak

\end{document}